%% file: main.tex
\documentclass[prc,twocolumn,floatfix,groupedaddress,nofootinbib,amsmath,amssymb,amsfonts,superscriptaddress,widetable] {revtex4-1}
\usepackage{bm}
\usepackage{mathrsfs}
\usepackage{amssymb}
\usepackage{amsmath}
\usepackage{graphicx}
\usepackage{relsize}
\usepackage{array}
\usepackage{xcolor}

\begin{document}

\title{Assessing the impact of uniform rotation on \\ the structure of neutron stars} 

\author{Marc Salinas and J. Piekarewicz}

\affiliation{Department of Physics, Florida State University, 
Tallahassee, FL 32306, USA}

\date{\today}
\begin{abstract}
Driven by recent laboratory experiments and astronomical observations, significant 
advances have deepened our understanding of neutron-star physics. NICER's Pulse 
Profile Modeling has refined our knowledge of neutron star masses and radii, while 
gravitational-wave detections have revealed key insights into the structure of neutron 
stars. Particularly relevant is the extraction of the tidal deformability by the LIGO-Virgo 
collaboration and the most recent determination of stellar radii by NICER, both 
suggesting a relatively soft equation of state (EOS) at intermediate densities.  Additionally, 
measurements from the PREX collaboration and from pulsar timing suggest instead that 
the EOS is stiff in the vicinity of saturation density and at the highest densities accessible 
to date. But how stiff can the EOS be at these very high densities? Recent events featuring 
compact objects near the ``lower mass gap" have raised questions about the existence of 
very massive neutron stars. Motivated by this finding and in light of new refinements to 
theoretical models, we explore the possibility that these massive objects may indeed be 
rapidly rotating neutron stars. We explore how rotation affects both the maximum neutron 
star mass and their associated radii, and discuss the implications they may have on the 
equation of state.
\end{abstract}
\smallskip

\maketitle

\section{Introduction}
\label{sec:introduction}
In the past decade, remarkable progress has been made in our understanding of the underlying 
dynamics of neutron stars. In terrestrial laboratories, the recent extraction of the neutron skin 
thickness of ${}^{208}$Pb by the PREX collaboration\,\cite{Adhikari:2021phr}  indicates that 
equation of state (EOS) in the vicinity of saturation density is stiff\,\cite{Reed:2021nqk}. 
In the cosmos, pulsar timing observations over long time periods have determined with high 
precision the individual masses of a binary system consisting of a white-dwarf star and the 
milisecond pulsar J0740+6620\,\cite{Cromartie:2019kug,Fonseca:2021wxt}. With a mass of 
$M\!=\!2.08\pm0.07\,M_{\odot}$, J0740 is one of the most massive neutron stars ever measured 
indicating that the equation of state must be stiff at the highest densities probed in the stellar 
interior. 

In turn, the Neutron Star Interior Composition Explorer (NICER) uses Pulse Profile Modeling 
to analyze X-ray emission from stellar hot spots, enabling the simultaneous determination of 
masses and radii. To date, four sources have been targeted:  
PSR J0030+0451\,\cite{Riley:2019yda,Miller:2019cac}, the previously mentioned millisecond 
pulsar PSR J0740+6620\,\cite{Riley:2021pdl,Miller:2021qha}, the brightest rotation-powered 
pulsar PSR J0437-4715\,\cite{Choudhury:2024xbk}, and PSR J1231-1411\,\cite{Salmi:2024bss}.
These pioneering observations---and future ones with enhanced sensitivity---are of critical 
importance given that it is possible to determine the nuclear EOS from the measurement 
of both masses and radii\,\cite{Lindblom:1992}. 

Finally, the historical detection of gravitational waves emitted from the binary coalescence of 
neutron stars (GW170817) has opened a brand new window into our 
Universe\,\cite{Abbott:PRL2017}. Concerning the EOS, the gravitational wave signal encodes
fundamental structural properties of the neutron star. The most precisely determined observable 
is the chirp mass, which involves a particular linear combination of the individual masses of the 
binary system. Of enormous relevance but significantly harder to extract, is the combined tidal 
deformability ($\tilde\Lambda$) a quantity that also involves a linear combination of the masses 
but now weighted by the individual tidal deformabilities of the two stars\,\cite{Abbott:PRL2017}. 
The tidal deformability encodes the linear response of the star to the tidal field generated by the 
companion and is an observable highly sensitive to the stellar compactness. In particular, the 
relatively small tidal deformability inferred from GW170817 suggests that both stars are compact 
and consequently that the EOS in the relevant density regime is 
soft\,\cite{Abbott:PRL2017,Abbott:2018exr}.

Given the significant progress achieved over the last few years, one may be on the verge of 
determining the maximum neutron star mass. Elucidating the maximum mass is a fundamental 
question that has implications in astrophysics, particle physics, and nuclear physics. From the 
perspective of particle physics, the maximum mass configuration determines the maximum density 
that can exist in the stellar interior, opening the possibility for the emergence of exotic states of 
matter, such as color superconductors\,\cite{Alford:1997zt,Alford:1998mk,Alford:2007xm}. In 
the context of nuclear science, knowledge of the maximum neutron star mass constrains the 
nuclear dynamics at the extremes of density and isospin asymmetry, conditions that are impossible 
to reproduce in terrestrial laboratories. Finally, in astrophysics the answer to this question sets the 
scale for the minimum black-hole mass, which illuminates, among other things, the origin of the 
intriguing ``lower mass gap" related to the absence of compact objects between three and five 
solar masses. Knowledge of the maximum neutron star mass also impacts the time-scale for the 
transition to a black hole during a binary neutron star coalescence which, in turn, affects the 
abundance of r-process elements. 

During the last five years, the Ligo-Virgo-Kagra (LVK) collaboration has detected gravitational waves 
from the coalescence of compact binaries with one of the components in, or near, the lower mass 
gap. The gravitational-wave event GW190814 involved the coalescence of a binary system with one 
of the most extreme mass ratios ever observed: a $23\,M_{\odot}$ black hole and a 
$2.59^{+0.08}_{-0.09}\,M_{\odot}$ ``compact" object tantalizingly near the lower mass 
gap\,\cite{Abbott:2020khf}. More recently, the LIGO-Virgo-Kagra (LVK) collaboration reported the 
coalescence of a compact binary---GW230529---with the primary component having a mass of 
$3.6^{+0.8}_{-1.2}\,M_{\odot}$\,\cite{Abac_2024}. In both cases, it has been suggested that the compact 
objects in (or near) the mass gap may be too heavy to be a neutron star\,\cite{Abbott:2020khf,Abac_2024}. 
Nevertheless, both papers left open the possibility that the compact object in the mass gap may be a neutron 
star, given that gravitational-data alone cannot exclude such possibility. We note that a recent electromagnetic 
observation of the pulsar binary PSR J0514-4002E places the (companion) mass of the compact object 
at a value of  2.09 and 2.71\,$M_{\odot}$\,\cite{Barr:2024wwl}, an interval that overlaps with the 
secondary object in GW190814.

Motivated by the discovery of GW190814---and given that the gravitational-wave signal by itself cannot 
exclude the existence of very massive neutron stars---an earlier publication explored\,\cite{Fattoyev:2020cws} 
the implications of a $2.6 M_{\odot}$ neutron star on a variety of observables sensitive to the stiffness of the 
equation of state. It was concluded at that time that for the set of covariant energy density functionals (EDFs)
explored in such a work, the stiffening of the EOS required to support super-massive neutron stars was found 
to be inconsistent with constraints obtained from both energetic heavy-ion collisions\,\cite{Danielewicz:2002pu} 
and from the low tidal deformability inferred from GW170817\,\cite{Abbott:PRL2017,Abbott:2018exr}. So if that 
was the case, what has changed since then that still motivates us to explore the possibility that the primary 
object in GW230529---an object even more massive than the secondary object in GW190814---may be a very 
massive neutron star?

We provide a three-prong response to the above question. First, a recent Bayesian analysis\,\cite{Oliinychenko:2022uvy} 
of flow data from the STAR collaboration seems to favor a hard equation of state at densities of about 2-3 nuclear matter 
saturation density. This is in stark contrast to the Danielewicz, Lacey, and Lynch analysis\,\cite{Danielewicz:2002pu} that 
ruled out strongly repulsive equations of state. Hence, the safest conclusion to draw from these competing studies is 
that at present there is no robust constraint from relativistic heavy-ion collisions on the high-density component of the 
nuclear EOS. Thus, on the basis of heavy-ion collisions alone, there is no longer any reason to reject the stiff equation 
of state proposed in our analysis of GW190814\,\cite{Fattoyev:2020cws}. Second, although the tidal deformability 
extracted from GW170817 suggests a soft EOS in the vicinity of twice saturation density\,\cite{Abbott:2018exr}, a re-analysis 
of the tidal parameters by Gamba and collaborators identified systematic errors in the waveform approximants as a major 
issue in the inference of the tidal parameters\,\cite{Gamba:2020wgg}. Note that the low tidal deformability suggested in the 
discovery paper\,\cite{Abbott:2018exr} was also used in Ref.\,\cite{Fattoyev:2020cws} as an argument against the existence
of very massive stars. Finally and importantly, in post-Newtonian theory there is a well known degeneracy between the mass 
ratio and the effective spin of the binary system. This implies that the extraction of the spin of the individual components 
will remain a significant challenge for some time to come. And whereas GW170817 suggests a low value for the effective 
spin parameter based on the notion that old neutron stars had ample time to spin down, this may not always be the case. 
Thus, if instead one assumes that the previously identified neutron stars may be spinning very fast, then it is not the mass 
of the non-rotating configuration that is relevant, but rather that of the rotating configuration. As such, we will assume 
throughout that the inferred masses of both the secondary object in GW190814 and the primary object in GW230529 may 
correspond to rapidly spinning neutron stars. Moreover, we note that NICER accounts for the rotation of the millisecond
pulsars by reporting the ``equatorial radius", a quantity that is often compared against theoretical predictions for static,
non-rotating neutron stars. Examining the impact of rotation on both the maximum mass configuration and stellar radii is 
the main goal of this work.

The manuscript has been organized as follows. In Sec.\ref{sec:formalism} we provide a brief review of the formalism 
used to compute the equation of state. Model uncertainties are quantified by adopting a few parameterizations that
generate both soft and stiff equations of state. In turn, these models serve as input into a TOV equation and into two
publicly available codes, RNS\,\cite{Stergioulas:1994ea,Stergioulas:2003yp} and LORENE, that will serve to assess 
the impact of uniform rotation. Our results are presented in Sec.\ref{sec:results} and we conclude by summarizing 
our results in Sec.\ref{sec:conclusions}.

\section{Formalism}
\label{sec:formalism}

As in much of our earlier work, covariant density functional theory will be the theoretical framework that will be employed to test 
the impact of rotation on the possible existence of massive neutron stars. In such a framework, a relativistic Lagrangian density 
consisting of nucleons and mesons is calibrated to the properties of finite nuclei and neutron stars. It is important to note that 
the equation of state EOS that serves as the sole input for the Tolman–Oppenheimer–Volkoff (TOV) equations is constructed 
from the same underlying Lagrangian density that  is used to compute the properties of finite nuclei, thereby connecting nuclear 
phenomena with length scales that differ by more than 18 orders of magnitude. 

The underlying Lagrangian density has been extensively reviewed in earlier publications\,\cite{Horowitz:2000xj,
Todd-Rutel:2005fa,Chen:2014sca,Yang:2019fvs}, so we limit ourselves to highlight the following two terms that are 
of particular importance to the high-density component of the equation of state:
\begin{equation}
{\mathscr L_{2}}\!=\frac{\zeta}{4!}  g_{\rm v}^4(V_{\mu}V^\mu)^2 +
                   \Lambda_{\rm v}\Big(g_{\rho}^{2}\,{\bf b}_{\mu}\cdot{\bf b}^{\mu}\Big)\!
                   \Big(g_{\rm v}^{2}V_{\nu}V^{\nu}\Big),
 \label{LDensity}
\end{equation}
where $g_{\rm v}$ and $g_{\rho}$ represent the strength of the Yukawa coupling of the nucleon to the 
isoscalar-vector ($V^{\mu}$) and isovector-vector (${\bf b}_{\mu}$) meson fields, respectively. Besides 
the standard Yukawa couplings, the Lagrangian density includes several meson self-interacting terms 
to account for density dependent effects. The first term in the above expression involves a quartic term 
in $V^{\mu}$ that was introduced by M\"uller and Serot to modify the high density component of the 
EOS\,\cite{Mueller:1996pm}. In particular, M\"uller and Serot showed that by tuning $\zeta$, one can 
change the maximum mass of a neutron star by up to one solar mass---with only minimum impact on 
the ground state properties of finite nuclei. In turn, the second term in the Lagrangian density induces 
mixing in the vector channel between the isoscalar and the isovector meson fields. The main motivation 
behind introducing $\Lambda_{\rm v}$ was to soften the symmetry energy---a quantity that plays a critical 
role in the determination of both the neutron skin thickness of neutron-rich nuclei and the radius of neutron 
stars\,\cite{Horowitz:2000xj}.  By properly tuning $\zeta$ and $\Lambda_{\rm v}$, one can stiffen the EOS 
of symmetric matter to generate massive neutron stars while softening the symmetry energy to obtain 
small stellar radii and tidal deformabilities. Indeed, it was precisely this approach that gave rise to the 
calibration of the ``BigApple" energy density functional, in the hope that one could account for the mass 
of the secondary object in GW190814\,\cite{Fattoyev:2020cws}. 

We close this section by listing in Table\,\ref{Table1} values for $\zeta$ and $\Lambda_{\rm v}$ for 
BigApple alongside two other energy density functionals: FSUGarnet\,\cite{Chen:2014mza} and 
FSUGold2\,\cite{Chen:2014sca}. The table displays predictions from these three models for stellar 
observables that are most sensitive to $\zeta$ and $\Lambda_{\rm v}$. For example, tuning $\zeta$ 
to small (non-negative) values stiffens the high density component of the EOS, resulting in large 
maximum masses $(M_{\rm TOV}$) as in the case of BigApple. In turn, increasing $\Lambda_{\rm v}$ 
softens the symmetry energy at the intermediate densities that control tidal deformabilities and 
stellar radii\cite{Lattimer:2006xb}. Such behavior is imprinted in the predictions of BigApple which 
generates both a large $(M_{\rm TOV}$) and moderate stellar radii. In contrast, both FSUGarnet
and FSUGold were calibrated to account for the $M_{\rm TOV}\!\gtrsim\!2\,M_{\odot}$ limit, but
with vastly different predictions for the stellar radius and tidal deformability of a $1.4\,M_{\odot}$ 
neutron star.

\begin{center}
\begin{table}[h]
\begin{tabular}{| l | | c | c | c | c | c |}
 \hline\rule{0pt}{2.5ex} 
\!\!Model  & $\zeta$ & $\Lambda_{\rm v}$  & $M_{\rm TOV}$ 
                & $R_{1.4}$ & $\Lambda_{1.4}$  \\
\hline
\hline
BigApple      & 0.00070  & 0.047471  &  2.600  & 12.960  &  717.3 \\
FSUGarnet  & 0.02350  & 0.043377  &  2.066  & 12.869  &  624.8 \\
FSUGold2   & 0.02560  & 0.000823  &  2.073  & 14.122  &  827.3  \\
\hline
\end{tabular}
\caption{The two model parameters $\zeta$ and $\Lambda_{\rm v}$, alongside
model predictions for the maximum mass of the static configuration (in solar masses), the radius (in km), and tidal deformability of a 1.4 solar mass neutron star.}
\label{Table1}
\end{table}
\end{center}

The results provided in Table\,\ref{Table1} were obtained by solving the Tolman-Oppenheimer-Volkoff equations 
for a static, non-rotating configuration. Uniform rotation---especially for rapid rotation near the mass shedding 
limit---requires the numerical solution of Einstein's equations. In the next chapter, two publicly available codes 
will be used to assess the impact of uniform rotation on the structure of neutron stars. The Rapidly Rotating 
Neutron Star (RNS) code is an efficient and easy to implement code designed to study rapidly rotating, relativistic, 
compact stars\,\cite{Stergioulas:1994ea,Stergioulas:2003yp}. In turn, LORENE is a software package developed 
to solve problems in numerical relativity using spectral methods\,\cite{Bonazzola:1998qx,Gourgoulhon:1999vx}. 
Both RNS and LORENE are flexible enough to allow for the input of generic equations of state, such as the ones
generated by the three covariant EDFs listed in Table\,\ref{Table1}. To ensure consistency, all results that will be 
presented in the following section were generated using both RNS and LORENE. 

\section{Results}
\label{sec:results}

\subsection{Impact of Uniform Rotation on Stellar Radii}
\label{sec:Radii}

\begin{figure*}[ht]
    \centering
    \includegraphics[width=0.95\textwidth]{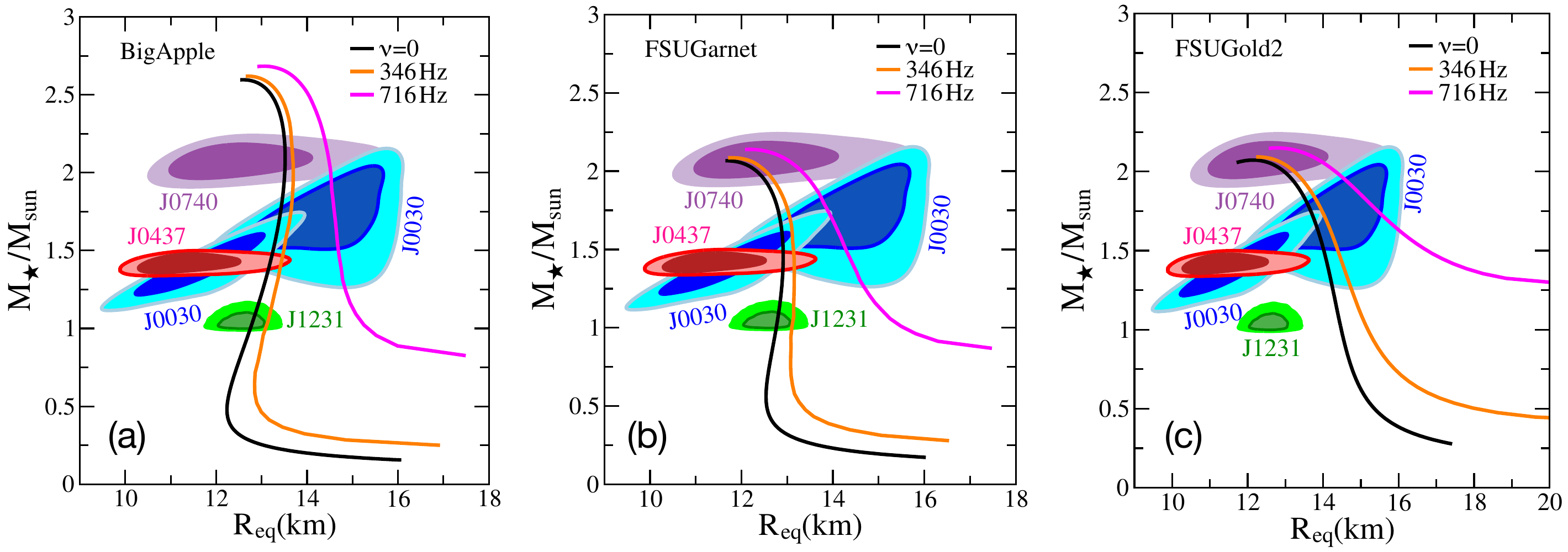}
    \caption{NICER 68\% and 95\% credible mass-radius intervals for their four target sources. Also
    shown are predictions from the three covariant energy density functionals introduced in the text;
    see Table\,\ref{Table1}. Predictions from these three models are made for the non-rotating configuration
    as well as for uniform rotation at the two frequencies listed on the inset.}
    \label{Figure1}
\end{figure*}

We start this section by assessing the impact of uniform rotation on four millisecond pulsars targeted by the 
NICER collaboration: PSR J0030+0451, PSR J0740+6620, PSR J0437-4715, and PSR J1231-1411, with 
rotational frequencies of 205 Hz\,\cite{Miller:2019cac}, 346\,Hz\,\cite{Cromartie:2019kug}, 
174\,Hz\,\cite{SosaFiscella:2020wmm}, and 271\,Hz\,\cite{Salmi:2024bss}, respectively. In Fig.\ref{Figure1} 
we display 68\% and 95\% credible Mass-Radius intervals for the four sources; for further details on the 
analysis see Ref.\cite{Chatziioannou:2024tjq} and references contained therein. Chronologically, PSR 
J0030+0451 was the first targeted source. However, relative to the initial 
publication\,\cite{Riley:2019yda,Miller:2019cac}, a recent reanalysis by Vinciguerra and 
collaborators\,\cite{Vinciguerra:2023qxq} using combined NICER and XMM-Newton data found a bimodal 
structure that is reflected in the two contours displayed in the figure. As such, J0030 does not impose any
significant constraint on the equation of state.

The next targeted source was PSR J0740+6620. Unlike PSR J0030+0451 where no mass measurement is 
currently available, an accurate determination of the mass of PSR J0740+6620 already 
existed\,\cite{Cromartie:2019kug,Fonseca:2021wxt}. This allowed for a determination of the stellar radius 
without any significant change in the mass from the prior pulsar timing observation. Note that the contours 
for PSR J0740+6620 are from Ref.\,\cite{Riley:2021pdl}, as Miller et al. reported a radius with a significantly 
larger central value\,\cite{Miller:2021qha}, although in both cases the uncertainties are large. Recently, 
NICER reported the mass and radius of PSR J0437-4715, the closest and brightest of the four observed
sources. Benefiting from available tight priors, Choudhury and collaborators inferred a mass largely unchanged 
from the prior value of $M\!=\!1.418\pm 0.044\,{\rm M}_{\odot}$\,\cite{Reardon:2024rdv}, but with a relatively 
small radius centered at $11.36\,{\rm km}$\,\cite{Choudhury:2024xbk}. Finally, we include in the figure the 
latest observation from the NICER mission---PSR J1231-1411, a millisecond pulsar with a mass in the 
neighborhood of one solar mass and an equatorial radius of about 12.6\,km\,\cite{Salmi:2024bss}. The
tight limit on the radius was obtained by demanding consistency with previous observational constraints 
and informed by chiral effective field theory. Instead, if the analysis used a largely uninformative prior, the 
equatorial radius increases by about one kilometer\,\cite{Salmi:2024bss}. 

To confront the NICER data, we have used the three models already introduced in Table\,\ref{Table1}
to compute mass-radius (MR) relations. Given that NICER reports equatorial radii for all four millisecond 
pulsars, we provide predictions for a non-rotating configuration ($\nu\!=\!0$) together with predictions 
for uniform rotation at $346$ and $716$\,Hz. We selected $346$\,Hz for being the highest frequency of 
the four NICER sources, while $716$\,Hz as the frequency of PSR J1748-2446ad, the fastest known 
spinning pulsar observed to date\,\cite{Hessels:2006ze}. In Figure\,\ref{Figure1}(a) we display predictions 
from BigApple, a covariant energy density functional constructed with the sole purpose of exploring the
possibility of generating very massive neutron stars\,\cite{Fattoyev:2020cws}. As seen in the figure, the 
model without rotation already predicts a maximum mass for the static configuration of 
$M_{\rm TOV}\!=\!2.6\,{\rm M}_{\odot}$; see also Table\,\ref{Table1}. As far as stellar radii, the model is
consistent with all four NICER sources, although falls within the 95\% contour for PSR J0437. Uniform 
rotation at $\nu\!=\!346$\,Hz has a negligible effects for the highest mass and only marginal impact at 
intermediate masses. Instead, neutron stars spinning at the highest frequency of $\nu\!=\!716$\,Hz, 
continue to have a minimal impact on the maximum mass but an appreciable effect on stellar radii, 
especially for low-mass stars. This is highly relevant given the relatively low mass of 
PSR J1231-1411\,\cite{Salmi:2024bss}. Although PSR J1231-1411 rotates at ``only" 271 Hz, we observe 
in Fig.\ref{Figure1}(a) that the equatorial radius of a putative low mass pulsar rotating at $716$\,Hz could 
increase by more than 2\,km.

The impact of uniform rotation on the other two models---FSUGarnet and FSUGold2---is qualitatively 
similar. Note that these two models were calibrated by demanding that they both satisfy the 2-solar 
mass constraint, but not more. The main difference between these two models is their prediction for the 
density dependence of the symmetry energy; FSUGarnet is soft whereas FSUGold2 is stiff. As such, 
FSUGold2 predicts both thick neutron skins and large stellar radii.  Although consistent with J0030 
and J0740, both J0437 and J1231 already rule out the large stellar radii predicted by FSUGold2. Note 
that for a model as stiff as FSUGold2, the shedding mass frequency is less than $716$Hz for neutron 
stars with masses below $1.3\,{\rm M}_{\odot}$. Interestingly, with a prediction of 
$R_{\rm skin}^{208}\!=\!0.287\pm\!0.020\,{\rm fm}$ for the neutron skin thickness of 
${}^{208}$Pb\,\cite{Chen:2014sca}, FSUGold2 is entirely consistent with the experimental value of 
$R_{\rm skin}^{208}\!=\!0.283\pm0.071\,{\rm fm}$ published by the PREX 
collaboration\,\cite{Adhikari:2021phr}. This could suggest a significant softening of the equation of
state between one and three times nuclear saturation density\,\cite{Reed:2021nqk}.

\subsection{Impact of Rotation on Maximum Masses}
\label{sec:Masses}
Whereas in the previous section we examined the impact of a ``modest" rotation on the mass-radius
relation, in this section we explore its impact on stellar masses at the mass shedding limit. 

We start with a simple exercise that assumes that realistic models, such as the ones that we have
introduced, can be pushed to such an extreme to generate maximum masses up to
$M_{\rm TOV}\!=\!2.8\,{\rm M}_{\odot}$, as demonstrated in Ref.\,\cite{Mueller:1996pm}. To assess 
the impact of maximum uniform rotation on the stability of the star, we evolve the static configuration 
by adopting a constant value for the ratio of masses, as suggested in the recent analysis of Musolino, 
Ecker, and Rezzolla\,\cite{Musolino:2023edi}; that is,

\begin{equation}
{\mathcal R} \equiv \frac{M_{\rm max}}{M_{\rm TOV}}=1.255^{+0.047}_{-0.040},
\label{MRatio}
\end{equation} 
where $M_{\rm TOV}$ is the maximum mass of the non-rotating (static) configuration and $M_{\rm max}$
is the maximum mass of the stable, uniformly rotating configuration. 

The probability distribution function (PDF) of the maximum mass configuration $M_{\rm max}$ may now be 
obtained from a ``product normal distribution" defined as\,\cite{Rohatgi:2015}:
\begin{equation}
P(M_{\rm max})\!=\!\int_{-\infty}^{\infty}\!\!\!P({\mathcal R})P(M_{\rm TOV})
                              \frac{dM_{\rm TOV}}{|M_{\rm TOV}|},
\label{PND}
\end{equation} 
where in the above expression ${\mathcal R}$ is the ratio of the two masses given in Eq.(\ref{MRatio}). Alternatively, 
one can generate the product normal distribution for $M_{\rm max}\!=\!{\mathcal R}\!\cdot\!M_{\rm TOV}$ by simply 
using Monte Carlo methods to sample from both $P({\mathcal R})$ and $P(M_{\rm TOV})$ distributions. 

However, a drastic simplification follows in the limit that the width of at least one of the two probability distributions 
is vanishingly small. For example, in the limit that the theoretical uncertainties in $M_{\rm TOV}$ can be 
ignored\,\cite{Chen:2014sca}, the product normal distribution reduces to a normal distribution with the following 
mean and standard deviations: $\mu_{\rm max}\!=\!\mu_{{\mathcal R}}\!\cdot\!\mu_{\rm TOV}$ and 
$\sigma_{\rm max}\!=\!\sigma_{\mathcal R}\!\cdot\!\mu_{\rm TOV}$. Mean values and standard deviations 
computed in this manner are listed in Table\,\ref{Table2}. 

\begin{table}[ht]
    \centering
    \begin{tabular}{|c|c||c|c|}
     \hline\rule{0pt}{2.5ex}  
       $M_{\rm TOV}$ & $M_{\rm max}$ &$\langle\Delta M_{\rm max}\rangle$ & $P_{{}_{\!\rm NS}}$ \\
       \hline
       \hline\rule{0pt}{2.5ex}  
    \!\!$2.20$ & $2.761(088)$  & $-0.811$  & $0.12$ \\
        $2.40$ & $3.012(096)$  &  $-0.562$  & $0.19$ \\
        $2.60$ & $3.263(104)$  & $-0.306$  & $0.28$ \\
        $2.80$ & $3.514(112)$  & $-0.057$ & $0.42$ \\
       \hline
    \end{tabular}
    \caption{Assumed maximum mass of the static configuration $M_{\rm TOV}$ and the resulting 
                 mean value and standard deviation for the mass of the maximally stable rotating 
                 configuration. The third column displays the average value of the 
                 $M_{\rm max}\!-\!M_{\rm LVK}$ distribution. The last column displays the probability that 
                 the primary object in GW230529 be a neutron star. All masses are in solar masses.}
    \label{Table2}
\end{table}

\begin{figure*}
    \centering
    \includegraphics[width=0.80\textwidth]{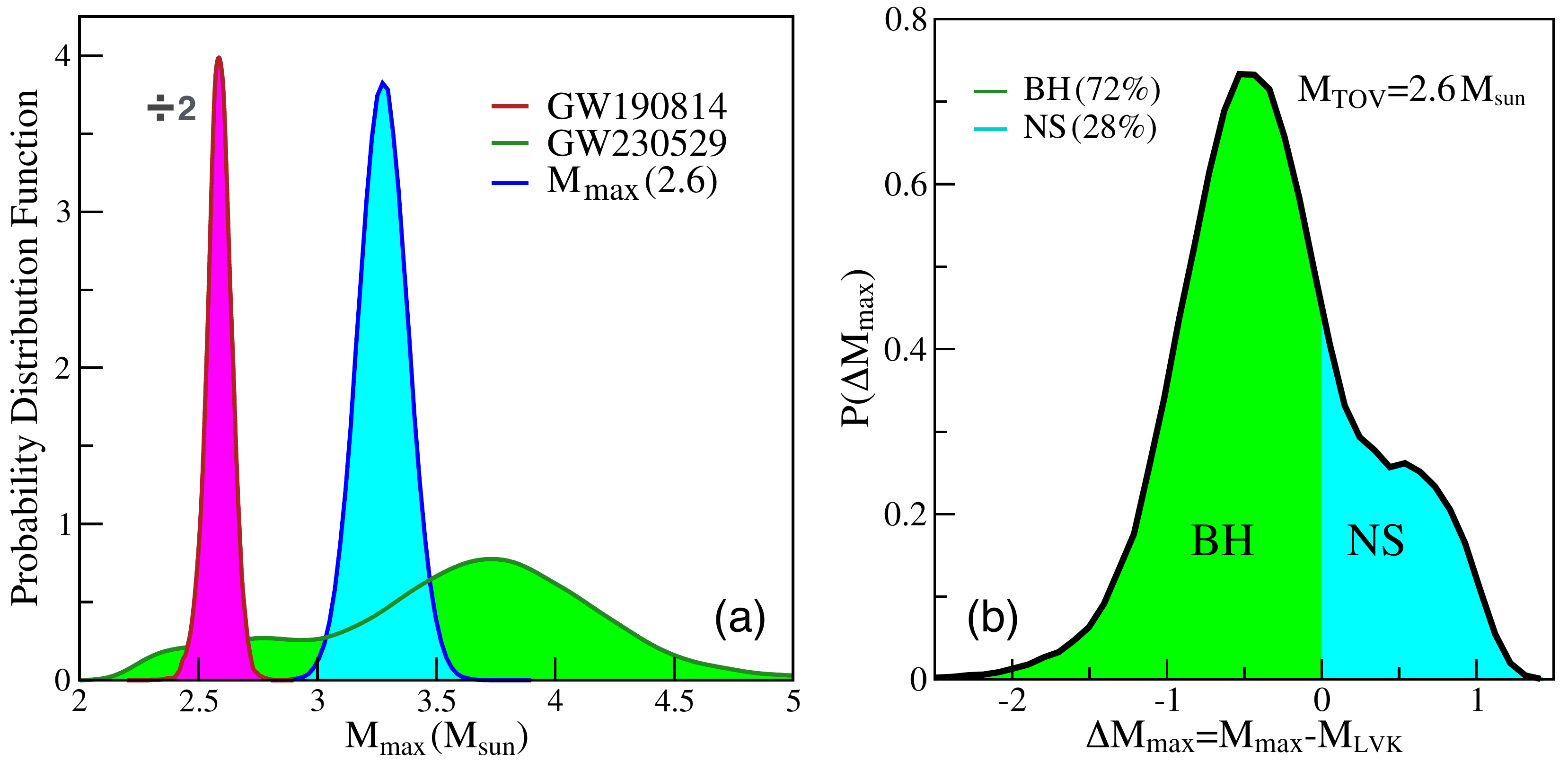}
    \caption{(a) Probability distribution functions for the secondary object in GW190814 (divided by 2)
                   and for the primary object in GW230529. Also shown is the impact of uniform rotation
                   at the mass shedding limit on a star with a TOV mass of $2.6\,M_{\odot}$, with the
                   rotational enhancement adopted from Eq.(\ref{MRatio}). (b) Probability distribution function 
                   for the difference between $M_{\rm max}$ and the mass of the primary object in GW230529.} 
    \label{Figure2}
\end{figure*}

Probability distribution functions are shown in Fig.\,\ref{Figure2}(a) for GW190814\,\cite{Abbott:2020khf},
for GW230529\,\cite{Abac_2024}, and for the maximum mass of the uniformly rotating configuration 
assuming a model that can support a static configuration with a maximum mass of 
$M_{\rm TOV}\!=\!2.6\,{\rm M}_{\odot}$, as in the case of BigApple. Note that the PDF for GW190814 
has been divided by a factor of 2.

Having generated the theoretical distribution of maximum masses, we can now generate a distribution 
for the mass difference between $M_{\rm max}$ and $M_{\rm LVK}$, where  $M_{\rm LVK}$ is either 
the secondary object in GW190814 or the primary object in GW230529. In this manner one frames 
the question of whether the enigmatic compact object is a neutron star as follows: what is the probability 
that the difference $\Delta M_{\rm max}\!\equiv\!M_{\rm max}\!-\!M_{\rm LVK}$ be positive? In the case 
of the secondary object in GW190814, the answer is evident from Fig.\,\ref{Figure2}(a), namely, the 
probability that the compact object is a neutron star is one. In the case of the primary object in GW230529, 
the answer is not evident because both the central value and the uncertainty are large. Hence, we display 
in Fig.\,\ref{Figure2}(b) the probability distribution $P(\Delta M_{\rm max})$. The area under the curve to the 
right of the origin represents the probability that the primary object in GW230529 be a neutron star; to the 
left that it is a black hole. Hence, under the assumed conditions, there is a 28\% probability that the 
primary object in GW230529 is a neutron star. Averages values for $P(\Delta M_{\rm max})$ and the 
probability that the primary object in GW230529 for various assumed values for $M_{\rm TOV}$ are 
listed in Table\,\ref{Table2}.

Although the schematic results displayed in Table\,\ref{Table2} and Fig.\,\ref{Figure2} provide credible 
estimates for the probability that the primary object in GW230529 be a neutron star, we now examine
the effects of rotation on the three covariant EDFs considered in this work. We list on the second and 
third columns of Table\,\ref{Table3} the maximum mass (in solar masses) and its associated radius 
(in km) for the non-rotating configuration. In turn, the last three columns display the Kepler frequency 
(in kHz) together with the mass and radius of the resulting maximally stable rotating configuration,
with the increase in the corresponding quantities shown in parenthesis. Whereas the $\sim\!20\%$ 
increase in ${\mathcal R}\!=\!M_{\rm max}/M_{\rm TOV}$ is consistent with the estimate from 
Ref.\,\cite{Musolino:2023edi}, it is interesting that the increase in the equatorial radius is even
larger. Clearly, any observable highly sensitive to the stellar compactness, such as the tidal
deformability, will be highly affected.

\begin{center}
\begin{table}[h]
\begin{tabular}{| l | | c | c || c | c | c |}
 \hline\rule{0pt}{2.5ex} 
\!\!Model  & $M_{\rm TOV}$ & $R_{\rm TOV}$ & 
  $\mathlarger{\mathlarger{\nu}}_{\rm K}$ 
                & $M_{\rm max}$ & $R_{\rm max}$  \\
\hline
\hline
BigApple      & 2.596 &  12.590 & 1.481 & 3.208(24\%)   &  16.727(33\%) \\
FSUGarnet  & 2.066  & 11.690  & 1.409 & 2.519(22\%)  & 16.048(37\%) \\
FSUGold2   & 2.073  & 12.151  & 1.321 & 2.463(19\%)  & 16.609(37\%)   \\
\hline
\end{tabular}
\caption{Maximum mass (in $M_{\odot}$) for the non-rotating and maximally stable 
rotating configurations and their corresponding equatorial radii (in km), with the 
Kepler frequency $\mathlarger{\mathlarger{\nu}}_{\rm K}$ given in kHz. The 
percentages denote the increase in mass and radius in going from the non-rotating 
to the rotating configuration.}
\label{Table3}
\end{table}
\end{center}

\begin{figure}[ht]
    \centering
    \includegraphics[width=0.45\textwidth]{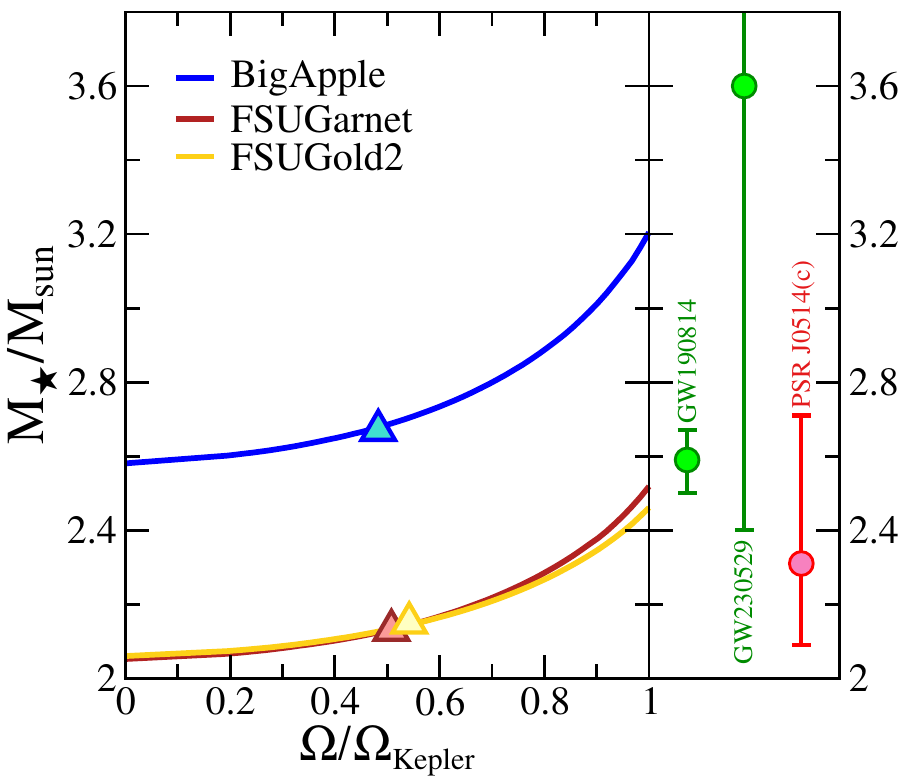}
    \caption{Evolution of the maximum mass as a function of rotational frequency 
             for the three covariant models considered in this paper. The triangles 
             denote rotation at the maximum observed limit of 716 Hz. The right-hand 
             panel displays central values and uncertainties for three heavy compact 
             objects that fall in (or near) the lower mass gap.}
    \label{Figure3}
\end{figure}

We conclude this section by examining the impact of uniform rotation on the structure 
of neutron stars away from the mass shedding limit. This is motivated by the fact that
at present there is no evidence that neutron stars could spin at---or even near---the 
mass shedding limit. For example, electromagnetic observations of  millisecond pulsars 
place the limit for maximum rotation at $716$\,Hz\,\cite{Hessels:2006ze}, a value that is 
significantly smaller than any of the Kepler frequencies listed in Table\,\ref{Table3}. Second, 
although difficult to determine, the spin of the individual components of coalescing neutron 
stars that merge over long period of times is expected to be low. Hence, we display
in Fig.\,\ref{Figure3} the evolution of the maximum mass as a function of frequency. Also 
shown are the masses for the secondary object in GW190814\,\cite{Abbott:2020khf},
the primary object in GW230529\,\cite{Abac_2024}, and for the companion mass of  
the millisecond pulsar PSR J0514\,\cite{Barr:2024wwl}. The figure shows a slow evolution 
at low frequencies that is then followed by a fairly rapid increase in the vicinity of the
Kepler limit. This is consistent with the results displayed in Fig.\ref{Figure1} that show 
no dramatic increase in the mass up to about half $\Omega_{\rm Kepler}$. However, 
we underscore that the impact of rotation on stellar radii could be significant. Finally,
we note that the predictions from BigApple at all frequencies---even in the static 
limit---are consistent with the three sources displayed in the figure, suggesting that
at this time it is difficult to rule out that the unidentified compact objects may be neutron
stars. 

\section{Conclusions}
\label{sec:conclusions}

The nature of matter at extreme temperatures and densities remains one of the most fundamental 
questions in nuclear astrophysics. Observations of massive neutron stars and their binary mergers 
play a critical role in answering this question. The determination of a neutron star's maximum mass 
and its associated radius provides powerful constraints on the maximum density achievable in its 
interior. In turn, this information serves as a cornerstone for validating the many fascinating phases 
that have been conjectured to exist in the stellar interior, such as quark and/or hyperonic matter.

In this work, we have examined available posterior distributions for the primary compact object in 
GW230529 and assessed its possibility of being a rapidly rotating neutron star. Under the framework
of covariant density functional theory, models such as BigApple---that are compatible with nuclear 
physics observables---place the upper limit on this probability at around 28\% for the maximally 
rotating configuration (see Fig.\ref{Figure2}). Despite this relatively low probability, our analysis suggests 
that the rapid rotation of neutron stars can substantially increase their maximum mass, opening the 
possibility that massive neutron stars, such as the one in GW230529, could indeed exist in the lower 
mass gap.

Moreover, our analysis on the uniform rotation of neutron stars seem to suggest that models consistent 
with nuclear physics may also be able to accommodate very heavy neutron stars---even in the absence 
of rotation. Indeed, as shown in Fig.\ref{Figure3}, the maximum mass predicted by BigApple for the 
non-rotating configuration accommodates the large mass of the secondary component of GW190814. 
This suggests that static configurations of neutron stars could also account for the existence of these 
massive compact objects, challenging previous assertions that such high masses necessarily imply 
black hole formation.

While our initial motivation was on how rotation affects neutron star masses, we discovered even more 
significant changes to stellar radii. This finding is particularly relevant to the NICER mission and future 
X-ray telescopes that target millisecond pulsars. For instance, our results revealed that the equatorial 
radius of a millisecond pulsar with a mass of approximately one solar mass rotating at the current maximum 
observed frequency, can vary by a few kilometers (see Fig.\ref{Figure1}).

Although significant uncertainties remain, particularly regarding the high density behavior of nuclear matter, 
our results affirm the potential for very massive neutron stars in the lower mass gap, both with and without 
rotation. These findings underscore the need for more precise constraints on the nuclear equation of state, 
especially at densities beyond saturation. Future experiments at the Facility for Rare Isotope Beams will
allow laboratory studies of asymmetric nuclear matter at twice normal nuclear density. Moreover, cosmic
observations, such as those from next-generation gravitational wave detectors and improved Pulse Profile 
Modeling, will be crucial in further constraining the mass-radius relation, assessing the impact of rotation 
on the structure of neutron stars, and validating the existence---or lack-thereof---of a lower mass gap.

\begin{acknowledgments}\vspace{-10pt}
We thank the ECT* and the FRIB Theory Alliance for support at the DTP/TALENT School on Nuclear 
Theory for Astrophysics during which this work was initiated. We would also like to thank our colleagues 
Devarshi Choudhury, Anna Puecher, Tuomo Salmi, and Serena Vinciguerra for many stimulating discussions,
and Prof. Frederi Viens for valuable insights into the statistical analysis. This material is based upon 
work supported by the U.S. Department of Energy Office of Science, Office of Nuclear Physics under Award 
Number DE-FG02-92ER40750. 
\end{acknowledgments}

\vfill\eject
\input{./main.bbl}

\end{document}

%% file: main.bbl
%

%% file: main.bbl
\begin{thebibliography}{40}%
\makeatletter
\providecommand \@ifxundefined [1]{%
 \@ifx{#1\undefined}
}%
\providecommand \@ifnum [1]{%
 \ifnum #1\expandafter \@firstoftwo
 \else \expandafter \@secondoftwo
 \fi
}%
\providecommand \@ifx [1]{%
 \ifx #1\expandafter \@firstoftwo
 \else \expandafter \@secondoftwo
 \fi
}%
\providecommand \natexlab [1]{#1}%
\providecommand \enquote  [1]{``#1''}%
\providecommand \bibnamefont  [1]{#1}%
\providecommand \bibfnamefont [1]{#1}%
\providecommand \citenamefont [1]{#1}%
\providecommand \href@noop [0]{\@secondoftwo}%
\providecommand \href [0]{\begingroup \@sanitize@url \@href}%
\providecommand \@href[1]{\@@startlink{#1}\@@href}%
\providecommand \@@href[1]{\endgroup#1\@@endlink}%
\providecommand \@sanitize@url [0]{\catcode `\\12\catcode `\$12\catcode
  `\&12\catcode `\#12\catcode `\^12\catcode `\_12\catcode `\%12\relax}%
\providecommand \@@startlink[1]{}%
\providecommand \@@endlink[0]{}%
\providecommand \url  [0]{\begingroup\@sanitize@url \@url }%
\providecommand \@url [1]{\endgroup\@href {#1}{\urlprefix }}%
\providecommand \urlprefix  [0]{URL }%
\providecommand \Eprint [0]{\href }%
\providecommand \doibase [0]{http://dx.doi.org/}%
\providecommand \selectlanguage [0]{\@gobble}%
\providecommand \bibinfo  [0]{\@secondoftwo}%
\providecommand \bibfield  [0]{\@secondoftwo}%
\providecommand \translation [1]{[#1]}%
\providecommand \BibitemOpen [0]{}%
\providecommand \bibitemStop [0]{}%
\providecommand \bibitemNoStop [0]{.\EOS\space}%
\providecommand \EOS [0]{\spacefactor3000\relax}%
\providecommand \BibitemShut  [1]{\csname bibitem#1\endcsname}%
\let\auto@bib@innerbib\@empty
\bibitem [{\citenamefont {Adhikari}\ \emph {et~al.}(2021)\citenamefont
  {Adhikari} \emph {et~al.}}]{Adhikari:2021phr}%
  \BibitemOpen
  \bibfield  {author} {\bibinfo {author} {\bibfnamefont {D.}~\bibnamefont
  {Adhikari}} \emph {et~al.} (\bibinfo {collaboration} {PREX}),\ }\href
  {\doibase 10.1103/PhysRevLett.126.172502} {\bibfield  {journal} {\bibinfo
  {journal} {Phys. Rev. Lett.}\ }\textbf {\bibinfo {volume} {126}},\ \bibinfo
  {pages} {172502} (\bibinfo {year} {2021})}\BibitemShut {NoStop}%
\bibitem [{\citenamefont {Reed}\ \emph {et~al.}(2021)\citenamefont {Reed},
  \citenamefont {Fattoyev}, \citenamefont {Horowitz},\ and\ \citenamefont
  {Piekarewicz}}]{Reed:2021nqk}%
  \BibitemOpen
  \bibfield  {author} {\bibinfo {author} {\bibfnamefont {B.~T.}\ \bibnamefont
  {Reed}}, \bibinfo {author} {\bibfnamefont {F.~J.}\ \bibnamefont {Fattoyev}},
  \bibinfo {author} {\bibfnamefont {C.~J.}\ \bibnamefont {Horowitz}}, \ and\
  \bibinfo {author} {\bibfnamefont {J.}~\bibnamefont {Piekarewicz}},\ }\href
  {\doibase 10.1103/PhysRevLett.126.172503} {\bibfield  {journal} {\bibinfo
  {journal} {Phys. Rev. Lett.}\ }\textbf {\bibinfo {volume} {126}},\ \bibinfo
  {pages} {172503} (\bibinfo {year} {2021})}\BibitemShut {NoStop}%
\bibitem [{\citenamefont {Cromartie}\ \emph {et~al.}(2019)\citenamefont
  {Cromartie} \emph {et~al.}}]{Cromartie:2019kug}%
  \BibitemOpen
  \bibfield  {author} {\bibinfo {author} {\bibfnamefont {H.~T.}\ \bibnamefont
  {Cromartie}} \emph {et~al.},\ }\href@noop {} {\bibfield  {journal} {\bibinfo
  {journal} {Nat. Astron.}\ }\textbf {\bibinfo {volume} {4}},\ \bibinfo {pages}
  {72} (\bibinfo {year} {2019})}\BibitemShut {NoStop}%
\bibitem [{\citenamefont {Fonseca}\ \emph {et~al.}(2021)\citenamefont {Fonseca}
  \emph {et~al.}}]{Fonseca:2021wxt}%
  \BibitemOpen
  \bibfield  {author} {\bibinfo {author} {\bibfnamefont {E.}~\bibnamefont
  {Fonseca}} \emph {et~al.},\ }\href {\doibase 10.3847/2041-8213/ac03b8}
  {\bibfield  {journal} {\bibinfo  {journal} {Astrophys. J. Lett.}\ }\textbf
  {\bibinfo {volume} {915}},\ \bibinfo {pages} {L12} (\bibinfo {year}
  {2021})}\BibitemShut {NoStop}%
\bibitem [{\citenamefont {Riley}\ \emph {et~al.}(2019)\citenamefont {Riley}
  \emph {et~al.}}]{Riley:2019yda}%
  \BibitemOpen
  \bibfield  {author} {\bibinfo {author} {\bibfnamefont {T.~E.}\ \bibnamefont
  {Riley}} \emph {et~al.},\ }\href {\doibase 10.3847/2041-8213/ab481c}
  {\bibfield  {journal} {\bibinfo  {journal} {Astrophys. J. Lett.}\ }\textbf
  {\bibinfo {volume} {887}},\ \bibinfo {pages} {L21} (\bibinfo {year}
  {2019})}\BibitemShut {NoStop}%
\bibitem [{\citenamefont {Miller}\ \emph {et~al.}(2019)\citenamefont {Miller}
  \emph {et~al.}}]{Miller:2019cac}%
  \BibitemOpen
  \bibfield  {author} {\bibinfo {author} {\bibfnamefont {M.~C.}\ \bibnamefont
  {Miller}} \emph {et~al.},\ }\href@noop {} {\bibfield  {journal} {\bibinfo
  {journal} {Astrophys. J. Lett.}\ }\textbf {\bibinfo {volume} {887}},\
  \bibinfo {pages} {L24} (\bibinfo {year} {2019})}\BibitemShut {NoStop}%
\bibitem [{\citenamefont {Riley}\ \emph {et~al.}(2021)\citenamefont {Riley}
  \emph {et~al.}}]{Riley:2021pdl}%
  \BibitemOpen
  \bibfield  {author} {\bibinfo {author} {\bibfnamefont {T.~E.}\ \bibnamefont
  {Riley}} \emph {et~al.},\ }\href {\doibase 10.3847/2041-8213/ac0a81}
  {\bibfield  {journal} {\bibinfo  {journal} {Astrophys. J. Lett.}\ }\textbf
  {\bibinfo {volume} {918}},\ \bibinfo {pages} {L27} (\bibinfo {year}
  {2021})}\BibitemShut {NoStop}%
\bibitem [{\citenamefont {Miller}\ \emph {et~al.}(2021)\citenamefont {Miller}
  \emph {et~al.}}]{Miller:2021qha}%
  \BibitemOpen
  \bibfield  {author} {\bibinfo {author} {\bibfnamefont {M.~C.}\ \bibnamefont
  {Miller}} \emph {et~al.},\ }\href {\doibase 10.3847/2041-8213/ac089b}
  {\bibfield  {journal} {\bibinfo  {journal} {Astrophys. J. Lett.}\ }\textbf
  {\bibinfo {volume} {918}},\ \bibinfo {pages} {L28} (\bibinfo {year}
  {2021})}\BibitemShut {NoStop}%
\bibitem [{\citenamefont {Choudhury}\ \emph {et~al.}(2024)\citenamefont
  {Choudhury} \emph {et~al.}}]{Choudhury:2024xbk}%
  \BibitemOpen
  \bibfield  {author} {\bibinfo {author} {\bibfnamefont {D.}~\bibnamefont
  {Choudhury}} \emph {et~al.},\ }\href {\doibase 10.3847/2041-8213/ad5a6f}
  {\bibfield  {journal} {\bibinfo  {journal} {Astrophys. J. Lett.}\ }\textbf
  {\bibinfo {volume} {971}},\ \bibinfo {pages} {L20} (\bibinfo {year}
  {2024})}\BibitemShut {NoStop}%
\bibitem [{\citenamefont {Salmi}\ \emph {et~al.}(2024)\citenamefont {Salmi}
  \emph {et~al.}}]{Salmi:2024bss}%
  \BibitemOpen
  \bibfield  {author} {\bibinfo {author} {\bibfnamefont {T.}~\bibnamefont
  {Salmi}} \emph {et~al.},\ }\href@noop {} {\  (\bibinfo {year} {2024})},\
  \Eprint {http://arxiv.org/abs/2409.14923} {arXiv:2409.14923 [astro-ph.HE]}
  \BibitemShut {NoStop}%
\bibitem [{\citenamefont {Lindblom}(1992)}]{Lindblom:1992}%
  \BibitemOpen
  \bibfield  {author} {\bibinfo {author} {\bibfnamefont {L.}~\bibnamefont
  {Lindblom}},\ }\href {\doibase 10.1086/171882} {\bibfield  {journal}
  {\bibinfo  {journal} {Astrophys. J.}\ }\textbf {\bibinfo {volume} {398}},\
  \bibinfo {pages} {569} (\bibinfo {year} {1992})}\BibitemShut {NoStop}%
\bibitem [{\citenamefont {Abbott}\ \emph {et~al.}(2017)\citenamefont {Abbott}
  \emph {et~al.}}]{Abbott:PRL2017}%
  \BibitemOpen
  \bibfield  {author} {\bibinfo {author} {\bibfnamefont {B.~P.}\ \bibnamefont
  {Abbott}} \emph {et~al.} (\bibinfo {collaboration} {Virgo, LIGO
  Scientific}),\ }\href {\doibase 10.1103/PhysRevLett.119.161101} {\bibfield
  {journal} {\bibinfo  {journal} {Phys. Rev. Lett.}\ }\textbf {\bibinfo
  {volume} {119}},\ \bibinfo {pages} {161101} (\bibinfo {year}
  {2017})}\BibitemShut {NoStop}%
\bibitem [{\citenamefont {Abbott}\ \emph {et~al.}(2018)\citenamefont {Abbott}
  \emph {et~al.}}]{Abbott:2018exr}%
  \BibitemOpen
  \bibfield  {author} {\bibinfo {author} {\bibfnamefont {B.~P.}\ \bibnamefont
  {Abbott}} \emph {et~al.} (\bibinfo {collaboration} {Virgo, LIGO
  Scientific}),\ }\href {\doibase 10.1103/PhysRevLett.121.161101} {\bibfield
  {journal} {\bibinfo  {journal} {Phys. Rev. Lett.}\ }\textbf {\bibinfo
  {volume} {121}},\ \bibinfo {pages} {161101} (\bibinfo {year}
  {2018})}\BibitemShut {NoStop}%
\bibitem [{\citenamefont {Alford}\ \emph {et~al.}(1998)\citenamefont {Alford},
  \citenamefont {Rajagopal},\ and\ \citenamefont {Wilczek}}]{Alford:1997zt}%
  \BibitemOpen
  \bibfield  {author} {\bibinfo {author} {\bibfnamefont {M.~G.}\ \bibnamefont
  {Alford}}, \bibinfo {author} {\bibfnamefont {K.}~\bibnamefont {Rajagopal}}, \
  and\ \bibinfo {author} {\bibfnamefont {F.}~\bibnamefont {Wilczek}},\
  }\href@noop {} {\bibfield  {journal} {\bibinfo  {journal} {Phys. Lett. B}\
  }\textbf {\bibinfo {volume} {422}},\ \bibinfo {pages} {247} (\bibinfo {year}
  {1998})}\BibitemShut {NoStop}%
\bibitem [{\citenamefont {Alford}\ \emph {et~al.}(1999)\citenamefont {Alford},
  \citenamefont {Rajagopal},\ and\ \citenamefont {Wilczek}}]{Alford:1998mk}%
  \BibitemOpen
  \bibfield  {author} {\bibinfo {author} {\bibfnamefont {M.~G.}\ \bibnamefont
  {Alford}}, \bibinfo {author} {\bibfnamefont {K.}~\bibnamefont {Rajagopal}}, \
  and\ \bibinfo {author} {\bibfnamefont {F.}~\bibnamefont {Wilczek}},\
  }\href@noop {} {\bibfield  {journal} {\bibinfo  {journal} {Nucl. Phys.}\
  }\textbf {\bibinfo {volume} {B537}} (\bibinfo {year} {1999})}\BibitemShut
  {NoStop}%
\bibitem [{\citenamefont {Alford}\ \emph {et~al.}(2008)\citenamefont {Alford},
  \citenamefont {Schmitt}, \citenamefont {Rajagopal},\ and\ \citenamefont
  {Schafer}}]{Alford:2007xm}%
  \BibitemOpen
  \bibfield  {author} {\bibinfo {author} {\bibfnamefont {M.~G.}\ \bibnamefont
  {Alford}}, \bibinfo {author} {\bibfnamefont {A.}~\bibnamefont {Schmitt}},
  \bibinfo {author} {\bibfnamefont {K.}~\bibnamefont {Rajagopal}}, \ and\
  \bibinfo {author} {\bibfnamefont {T.}~\bibnamefont {Schafer}},\ }\href
  {\doibase 10.1103/RevModPhys.80.1455} {\bibfield  {journal} {\bibinfo
  {journal} {Rev. Mod. Phys.}\ }\textbf {\bibinfo {volume} {80}},\ \bibinfo
  {pages} {1455} (\bibinfo {year} {2008})}\BibitemShut {NoStop}%
\bibitem [{\citenamefont {Abbott}\ \emph {et~al.}(2020)\citenamefont {Abbott}
  \emph {et~al.}}]{Abbott:2020khf}%
  \BibitemOpen
  \bibfield  {author} {\bibinfo {author} {\bibfnamefont {R.}~\bibnamefont
  {Abbott}} \emph {et~al.} (\bibinfo {collaboration} {LIGO-Virgo
  Collaboration}),\ }\href {\doibase 10.3847/2041-8213/ab960f} {\bibfield
  {journal} {\bibinfo  {journal} {Astrophys. J.}\ }\textbf {\bibinfo {volume}
  {896}},\ \bibinfo {pages} {L44} (\bibinfo {year} {2020})}\BibitemShut
  {NoStop}%
\bibitem [{\citenamefont {Abac}\ \emph {et~al.}(2024)\citenamefont {Abac} \emph
  {et~al.}}]{Abac_2024}%
  \BibitemOpen
  \bibfield  {author} {\bibinfo {author} {\bibfnamefont {A.~G.}\ \bibnamefont
  {Abac}} \emph {et~al.} (\bibinfo {collaboration} {The LIGO Scientific
  Collaboration, the Virgo Collaboration, and the KAGRA Collaboration}),\
  }\href {\doibase 10.3847/2041-8213/ad5beb} {\bibfield  {journal} {\bibinfo
  {journal} {Astrophys. J.}\ }\textbf {\bibinfo {volume} {970}},\ \bibinfo
  {pages} {L34} (\bibinfo {year} {2024})}\BibitemShut {NoStop}%
\bibitem [{\citenamefont {Barr}\ \emph {et~al.}(2024)\citenamefont {Barr} \emph
  {et~al.}}]{Barr:2024wwl}%
  \BibitemOpen
  \bibfield  {author} {\bibinfo {author} {\bibfnamefont {E.~D.}\ \bibnamefont
  {Barr}} \emph {et~al.},\ }\href {\doibase 10.1126/science.adg3005} {\bibfield
   {journal} {\bibinfo  {journal} {Science}\ }\textbf {\bibinfo {volume}
  {383}},\ \bibinfo {pages} {275} (\bibinfo {year} {2024})}\BibitemShut
  {NoStop}%
\bibitem [{\citenamefont {Fattoyev}\ \emph {et~al.}(2020)\citenamefont
  {Fattoyev}, \citenamefont {Horowitz}, \citenamefont {Piekarewicz},\ and\
  \citenamefont {Reed}}]{Fattoyev:2020cws}%
  \BibitemOpen
  \bibfield  {author} {\bibinfo {author} {\bibfnamefont {F.~J.}\ \bibnamefont
  {Fattoyev}}, \bibinfo {author} {\bibfnamefont {C.~J.}\ \bibnamefont
  {Horowitz}}, \bibinfo {author} {\bibfnamefont {J.}~\bibnamefont
  {Piekarewicz}}, \ and\ \bibinfo {author} {\bibfnamefont {B.}~\bibnamefont
  {Reed}},\ }\href {\doibase 10.1103/PhysRevC.102.065805} {\bibfield  {journal}
  {\bibinfo  {journal} {Phys. Rev. C}\ }\textbf {\bibinfo {volume} {102}},\
  \bibinfo {pages} {065805} (\bibinfo {year} {2020})}\BibitemShut {NoStop}%
\bibitem [{\citenamefont {Danielewicz}\ \emph {et~al.}(2002)\citenamefont
  {Danielewicz}, \citenamefont {Lacey},\ and\ \citenamefont
  {Lynch}}]{Danielewicz:2002pu}%
  \BibitemOpen
  \bibfield  {author} {\bibinfo {author} {\bibfnamefont {P.}~\bibnamefont
  {Danielewicz}}, \bibinfo {author} {\bibfnamefont {R.}~\bibnamefont {Lacey}},
  \ and\ \bibinfo {author} {\bibfnamefont {W.~G.}\ \bibnamefont {Lynch}},\
  }\href@noop {} {\bibfield  {journal} {\bibinfo  {journal} {Science}\ }\textbf
  {\bibinfo {volume} {298}},\ \bibinfo {pages} {1592} (\bibinfo {year}
  {2002})}\BibitemShut {NoStop}%
\bibitem [{\citenamefont {Oliinychenko}\ \emph {et~al.}(2023)\citenamefont
  {Oliinychenko}, \citenamefont {Sorensen}, \citenamefont {Koch},\ and\
  \citenamefont {McLerran}}]{Oliinychenko:2022uvy}%
  \BibitemOpen
  \bibfield  {author} {\bibinfo {author} {\bibfnamefont {D.}~\bibnamefont
  {Oliinychenko}}, \bibinfo {author} {\bibfnamefont {A.}~\bibnamefont
  {Sorensen}}, \bibinfo {author} {\bibfnamefont {V.}~\bibnamefont {Koch}}, \
  and\ \bibinfo {author} {\bibfnamefont {L.}~\bibnamefont {McLerran}},\ }\href
  {\doibase 10.1103/PhysRevC.108.034908} {\bibfield  {journal} {\bibinfo
  {journal} {Phys. Rev. C}\ }\textbf {\bibinfo {volume} {108}},\ \bibinfo
  {pages} {034908} (\bibinfo {year} {2023})}\BibitemShut {NoStop}%
\bibitem [{\citenamefont {Gamba}\ \emph {et~al.}(2021)\citenamefont {Gamba},
  \citenamefont {Breschi}, \citenamefont {Bernuzzi}, \citenamefont {Agathos},\
  and\ \citenamefont {Nagar}}]{Gamba:2020wgg}%
  \BibitemOpen
  \bibfield  {author} {\bibinfo {author} {\bibfnamefont {R.}~\bibnamefont
  {Gamba}}, \bibinfo {author} {\bibfnamefont {M.}~\bibnamefont {Breschi}},
  \bibinfo {author} {\bibfnamefont {S.}~\bibnamefont {Bernuzzi}}, \bibinfo
  {author} {\bibfnamefont {M.}~\bibnamefont {Agathos}}, \ and\ \bibinfo
  {author} {\bibfnamefont {A.}~\bibnamefont {Nagar}},\ }\href {\doibase
  10.1103/PhysRevD.103.124015} {\bibfield  {journal} {\bibinfo  {journal}
  {Phys. Rev. D}\ }\textbf {\bibinfo {volume} {103}},\ \bibinfo {pages}
  {124015} (\bibinfo {year} {2021})}\BibitemShut {NoStop}%
\bibitem [{\citenamefont {Stergioulas}\ and\ \citenamefont
  {Friedman}(1995)}]{Stergioulas:1994ea}%
  \BibitemOpen
  \bibfield  {author} {\bibinfo {author} {\bibfnamefont {N.}~\bibnamefont
  {Stergioulas}}\ and\ \bibinfo {author} {\bibfnamefont {J.~L.}\ \bibnamefont
  {Friedman}},\ }\href {\doibase 10.1086/175605} {\bibfield  {journal}
  {\bibinfo  {journal} {Astrophys. J.}\ }\textbf {\bibinfo {volume} {444}},\
  \bibinfo {pages} {306} (\bibinfo {year} {1995})}\BibitemShut {NoStop}%
\bibitem [{\citenamefont {Stergioulas}(2003)}]{Stergioulas:2003yp}%
  \BibitemOpen
  \bibfield  {author} {\bibinfo {author} {\bibfnamefont {N.}~\bibnamefont
  {Stergioulas}},\ }\href {\doibase 10.12942/lrr-2003-3} {\bibfield  {journal}
  {\bibinfo  {journal} {Living Rev. Rel.}\ }\textbf {\bibinfo {volume} {6}},\
  \bibinfo {pages} {3} (\bibinfo {year} {2003})}\BibitemShut {NoStop}%
\bibitem [{\citenamefont {Horowitz}\ and\ \citenamefont
  {Piekarewicz}(2001)}]{Horowitz:2000xj}%
  \BibitemOpen
  \bibfield  {author} {\bibinfo {author} {\bibfnamefont {C.~J.}\ \bibnamefont
  {Horowitz}}\ and\ \bibinfo {author} {\bibfnamefont {J.}~\bibnamefont
  {Piekarewicz}},\ }\href@noop {} {\bibfield  {journal} {\bibinfo  {journal}
  {Phys. Rev. Lett.}\ }\textbf {\bibinfo {volume} {86}},\ \bibinfo {pages}
  {5647} (\bibinfo {year} {2001})}\BibitemShut {NoStop}%
\bibitem [{\citenamefont {Todd-Rutel}\ and\ \citenamefont
  {Piekarewicz}(2005)}]{Todd-Rutel:2005fa}%
  \BibitemOpen
  \bibfield  {author} {\bibinfo {author} {\bibfnamefont {B.~G.}\ \bibnamefont
  {Todd-Rutel}}\ and\ \bibinfo {author} {\bibfnamefont {J.}~\bibnamefont
  {Piekarewicz}},\ }\href@noop {} {\bibfield  {journal} {\bibinfo  {journal}
  {Phys. Rev. Lett}\ }\textbf {\bibinfo {volume} {95}},\ \bibinfo {pages}
  {122501} (\bibinfo {year} {2005})}\BibitemShut {NoStop}%
\bibitem [{\citenamefont {Chen}\ and\ \citenamefont
  {Piekarewicz}(2014)}]{Chen:2014sca}%
  \BibitemOpen
  \bibfield  {author} {\bibinfo {author} {\bibfnamefont {W.-C.}\ \bibnamefont
  {Chen}}\ and\ \bibinfo {author} {\bibfnamefont {J.}~\bibnamefont
  {Piekarewicz}},\ }\href@noop {} {\bibfield  {journal} {\bibinfo  {journal}
  {Phys. Rev.}\ }\textbf {\bibinfo {volume} {C90}},\ \bibinfo {pages} {044305}
  (\bibinfo {year} {2014})}\BibitemShut {NoStop}%
\bibitem [{\citenamefont {Yang}\ and\ \citenamefont
  {Piekarewicz}(2020)}]{Yang:2019fvs}%
  \BibitemOpen
  \bibfield  {author} {\bibinfo {author} {\bibfnamefont {J.}~\bibnamefont
  {Yang}}\ and\ \bibinfo {author} {\bibfnamefont {J.}~\bibnamefont
  {Piekarewicz}},\ }\href {\doibase 10.1146/annurev-nucl-101918-023608}
  {\bibfield  {journal} {\bibinfo  {journal} {Ann. Rev. Nucl. Part. Sci.}\
  }\textbf {\bibinfo {volume} {70}},\ \bibinfo {pages} {21} (\bibinfo {year}
  {2020})}\BibitemShut {NoStop}%
\bibitem [{\citenamefont {Mueller}\ and\ \citenamefont
  {Serot}(1996)}]{Mueller:1996pm}%
  \BibitemOpen
  \bibfield  {author} {\bibinfo {author} {\bibfnamefont {H.}~\bibnamefont
  {Mueller}}\ and\ \bibinfo {author} {\bibfnamefont {B.~D.}\ \bibnamefont
  {Serot}},\ }\href@noop {} {\bibfield  {journal} {\bibinfo  {journal} {Nucl.
  Phys.}\ }\textbf {\bibinfo {volume} {A606}},\ \bibinfo {pages} {508}
  (\bibinfo {year} {1996})}\BibitemShut {NoStop}%
\bibitem [{\citenamefont {Chen}\ and\ \citenamefont
  {Piekarewicz}(2015)}]{Chen:2014mza}%
  \BibitemOpen
  \bibfield  {author} {\bibinfo {author} {\bibfnamefont {W.-C.}\ \bibnamefont
  {Chen}}\ and\ \bibinfo {author} {\bibfnamefont {J.}~\bibnamefont
  {Piekarewicz}},\ }\href@noop {} {\bibfield  {journal} {\bibinfo  {journal}
  {Phys. Lett.}\ }\textbf {\bibinfo {volume} {B748}},\ \bibinfo {pages} {284}
  (\bibinfo {year} {2015})}\BibitemShut {NoStop}%
\bibitem [{\citenamefont {Lattimer}\ and\ \citenamefont
  {Prakash}(2007)}]{Lattimer:2006xb}%
  \BibitemOpen
  \bibfield  {author} {\bibinfo {author} {\bibfnamefont {J.~M.}\ \bibnamefont
  {Lattimer}}\ and\ \bibinfo {author} {\bibfnamefont {M.}~\bibnamefont
  {Prakash}},\ }\href@noop {} {\bibfield  {journal} {\bibinfo  {journal} {Phys.
  Rept.}\ }\textbf {\bibinfo {volume} {442}},\ \bibinfo {pages} {109} (\bibinfo
  {year} {2007})}\BibitemShut {NoStop}%
\bibitem [{\citenamefont {Bonazzola}\ \emph {et~al.}(1998)\citenamefont
  {Bonazzola}, \citenamefont {Gourgoulhon},\ and\ \citenamefont
  {Marck}}]{Bonazzola:1998qx}%
  \BibitemOpen
  \bibfield  {author} {\bibinfo {author} {\bibfnamefont {S.}~\bibnamefont
  {Bonazzola}}, \bibinfo {author} {\bibfnamefont {E.}~\bibnamefont
  {Gourgoulhon}}, \ and\ \bibinfo {author} {\bibfnamefont {J.-A.}\ \bibnamefont
  {Marck}},\ }\href {\doibase 10.1103/PhysRevD.58.104020} {\bibfield  {journal}
  {\bibinfo  {journal} {Phys. Rev. D}\ }\textbf {\bibinfo {volume} {58}},\
  \bibinfo {pages} {104020} (\bibinfo {year} {1998})}\BibitemShut {NoStop}%
\bibitem [{\citenamefont {Gourgoulhon}\ \emph {et~al.}(1999)\citenamefont
  {Gourgoulhon}, \citenamefont {Haensel}, \citenamefont {Livine}, \citenamefont
  {Paluch}, \citenamefont {Bonazzola},\ and\ \citenamefont
  {Marck}}]{Gourgoulhon:1999vx}%
  \BibitemOpen
  \bibfield  {author} {\bibinfo {author} {\bibfnamefont {E.}~\bibnamefont
  {Gourgoulhon}}, \bibinfo {author} {\bibfnamefont {P.}~\bibnamefont
  {Haensel}}, \bibinfo {author} {\bibfnamefont {R.}~\bibnamefont {Livine}},
  \bibinfo {author} {\bibfnamefont {E.}~\bibnamefont {Paluch}}, \bibinfo
  {author} {\bibfnamefont {S.}~\bibnamefont {Bonazzola}}, \ and\ \bibinfo
  {author} {\bibfnamefont {J.~A.}\ \bibnamefont {Marck}},\ }\href@noop {}
  {\bibfield  {journal} {\bibinfo  {journal} {Astron. Astrophys.}\ }\textbf
  {\bibinfo {volume} {349}},\ \bibinfo {pages} {851} (\bibinfo {year}
  {1999})}\BibitemShut {NoStop}%
\bibitem [{\citenamefont {Sosa~Fiscella}\ \emph {et~al.}(2021)\citenamefont
  {Sosa~Fiscella} \emph {et~al.}}]{SosaFiscella:2020wmm}%
  \BibitemOpen
  \bibfield  {author} {\bibinfo {author} {\bibfnamefont {V.}~\bibnamefont
  {Sosa~Fiscella}} \emph {et~al.},\ }\href {\doibase 10.3847/1538-4357/abceb3}
  {\bibfield  {journal} {\bibinfo  {journal} {Astrophys. J.}\ }\textbf
  {\bibinfo {volume} {908}},\ \bibinfo {pages} {158} (\bibinfo {year}
  {2021})}\BibitemShut {NoStop}%
\bibitem [{\citenamefont {Chatziioannou}\ \emph {et~al.}(2024)\citenamefont
  {Chatziioannou}, \citenamefont {Cromartie}, \citenamefont {Gandolfi},
  \citenamefont {Tews}, \citenamefont {Radice}, \citenamefont {Steiner},\ and\
  \citenamefont {Watts}}]{Chatziioannou:2024tjq}%
  \BibitemOpen
  \bibfield  {author} {\bibinfo {author} {\bibfnamefont {K.}~\bibnamefont
  {Chatziioannou}}, \bibinfo {author} {\bibfnamefont {H.~T.}\ \bibnamefont
  {Cromartie}}, \bibinfo {author} {\bibfnamefont {S.}~\bibnamefont {Gandolfi}},
  \bibinfo {author} {\bibfnamefont {I.}~\bibnamefont {Tews}}, \bibinfo {author}
  {\bibfnamefont {D.}~\bibnamefont {Radice}}, \bibinfo {author} {\bibfnamefont
  {A.~W.}\ \bibnamefont {Steiner}}, \ and\ \bibinfo {author} {\bibfnamefont
  {A.~L.}\ \bibnamefont {Watts}},\ }\href@noop {} {\  (\bibinfo {year}
  {2024})},\ \Eprint {http://arxiv.org/abs/2407.11153} {arXiv:2407.11153
  [nucl-th]} \BibitemShut {NoStop}%
\bibitem [{\citenamefont {Vinciguerra}\ \emph {et~al.}(2024)\citenamefont
  {Vinciguerra} \emph {et~al.}}]{Vinciguerra:2023qxq}%
  \BibitemOpen
  \bibfield  {author} {\bibinfo {author} {\bibfnamefont {S.}~\bibnamefont
  {Vinciguerra}} \emph {et~al.},\ }\href {\doibase 10.3847/1538-4357/acfb83}
  {\bibfield  {journal} {\bibinfo  {journal} {Astrophys. J.}\ }\textbf
  {\bibinfo {volume} {961}},\ \bibinfo {pages} {62} (\bibinfo {year}
  {2024})}\BibitemShut {NoStop}%
\bibitem [{\citenamefont {Reardon}\ \emph {et~al.}(2024)\citenamefont {Reardon}
  \emph {et~al.}}]{Reardon:2024rdv}%
  \BibitemOpen
  \bibfield  {author} {\bibinfo {author} {\bibfnamefont {D.~J.}\ \bibnamefont
  {Reardon}} \emph {et~al.},\ }\href {\doibase 10.3847/2041-8213/ad614a}
  {\bibfield  {journal} {\bibinfo  {journal} {Astrophys. J. Lett.}\ }\textbf
  {\bibinfo {volume} {971}},\ \bibinfo {pages} {L18} (\bibinfo {year}
  {2024})}\BibitemShut {NoStop}%
\bibitem [{\citenamefont {Hessels}\ \emph {et~al.}(2006)\citenamefont
  {Hessels}, \citenamefont {Ransom}, \citenamefont {Stairs}, \citenamefont
  {Freire}, \citenamefont {Kaspi},\ and\ \citenamefont
  {Camilo}}]{Hessels:2006ze}%
  \BibitemOpen
  \bibfield  {author} {\bibinfo {author} {\bibfnamefont {J.~W.~T.}\
  \bibnamefont {Hessels}}, \bibinfo {author} {\bibfnamefont {S.~M.}\
  \bibnamefont {Ransom}}, \bibinfo {author} {\bibfnamefont {I.~H.}\
  \bibnamefont {Stairs}}, \bibinfo {author} {\bibfnamefont {P.~C.~C.}\
  \bibnamefont {Freire}}, \bibinfo {author} {\bibfnamefont {V.~M.}\
  \bibnamefont {Kaspi}}, \ and\ \bibinfo {author} {\bibfnamefont
  {F.}~\bibnamefont {Camilo}},\ }\href {\doibase 10.1126/science.1123430}
  {\bibfield  {journal} {\bibinfo  {journal} {Science}\ }\textbf {\bibinfo
  {volume} {311}},\ \bibinfo {pages} {1901} (\bibinfo {year}
  {2006})}\BibitemShut {NoStop}%
\bibitem [{\citenamefont {Musolino}\ \emph {et~al.}(2024)\citenamefont
  {Musolino}, \citenamefont {Ecker},\ and\ \citenamefont
  {Rezzolla}}]{Musolino:2023edi}%
  \BibitemOpen
  \bibfield  {author} {\bibinfo {author} {\bibfnamefont {C.}~\bibnamefont
  {Musolino}}, \bibinfo {author} {\bibfnamefont {C.}~\bibnamefont {Ecker}}, \
  and\ \bibinfo {author} {\bibfnamefont {L.}~\bibnamefont {Rezzolla}},\ }\href
  {\doibase 10.3847/1538-4357/ad1758} {\bibfield  {journal} {\bibinfo
  {journal} {Astrophys. J.}\ }\textbf {\bibinfo {volume} {962}},\ \bibinfo
  {pages} {61} (\bibinfo {year} {2024})}\BibitemShut {NoStop}%
\end{thebibliography}
